\documentclass[aip,apl,10pt,floatfix,twocolumn]{revtex4-1}  

\usepackage{graphicx}  
\usepackage{amssymb}   

\usepackage[english]{babel}

\newcommand{\JQI}{Joint Quantum Institute, National Institute of Standards and Technology, \& University of Maryland, College Park, MD, USA}
\newcommand{\NIST}{National Institute of Standards and Technology, Gaithersburg, MD, USA}
\newcommand{\un}[1]{\ensuremath{\;\mathrm{#1}}}
\newcommand{\g}{\ensuremath{g^{(2)}(0)}}

\begin{document}\emph{}

\title{Filter-free single-photon quantum dot resonance fluorescence in an integrated cavity-waveguide device}
\author{Tobias Huber}
\altaffiliation{Current address: Technische Physik, Universit\"{a}at W\"{u}rzburg, W\"{u}rzburg (tobias.j.huber@gmail.com)} 
\affiliation{\JQI}
\author{Marcelo Davan\c{c}o}
\affiliation{\NIST}
\author{Markus M{\"u}ller}
\affiliation{\JQI}
\author{Yichen Shuai}
\affiliation{\JQI}
\author{Olivier Gazzano}
\affiliation{\JQI}
\author{Glenn S. Solomon}
\email{gsolomon@umd.edu}
\affiliation{\JQI}
\affiliation{\NIST}
\date{\today}

\begin{abstract}
Semiconductor quantum dots embedded in micro-pillar cavities are excellent emitters of  single photons when pumped resonantly. Often, the same spatial mode is used to both resonantly excite a quantum dot and to collect the emitted single photons, requiring cross-polarization to reduce the uncoupled scattered laser light. This inherently reduces the source brightness to 50~\%. Critically, for some quantum applications the total efficiency from generation to detection must be over 50~\%. Here, we demonstrate a resonant-excitation approach to creating single photons that is free of any cross-polarization, and in fact any filtering whatsoever. It potentially increases single-photon rates and collection efficiencies, and simplifies operation.
 This integrated device allows us to resonantly excite single quantum-dot states in several cavities in the plane of the device using connected waveguides, while the cavity-enhanced single-photon fluorescence is directed vertical (off-chip) in a Gaussian mode. We expect this design to be a prototype for larger chip-scale quantum photonics.
\end{abstract}

\maketitle

In the on-going development of quantum optical technologies, devices will need to be easier to use, more compact, robust, and scalable, making them available to a broader community. These technologies include applications in quantum communication~\citep{Gisin2007, Simon2007,Ursin2007,Liao2017}, optical quantum metrology~\citep{Giovannetti2004,Banaszek2009,Matthews2016,muller2017}, and optical quantum computation and simulation~\citep{Knill2001, Kok2007, OBrien2007, Gazzano2016}. For example, a true single photon source on chip as a turnkey device would open quantum technologies to a unprecedented user group. 

Quantum dot (QD) excitonic states are excellent quantum emitters, showing bright emission of
single photons~\citep{Michler2000, Solomon2001, Gazzano2013, Somaschi2016, Senellart2017} and excellent suppression of multi-photons states~\citep{Jayakumar2013,Somaschi2016,Schweickert2018}. These properties are achieved due to the level structure and radiative efficiency of the optically allowed lowest level exciton states.

While single QD exciton emission is inherently bright with low multi-photon contribution, the emitted light can be further enhanced and directed into a Gaussian mode by coupling the QD to an optical cavity.~\citep{Haroche1989,Gerard1998, Solomon2001} 
In the weak coupling regime between emitter and cavity, this is known as the Purcell effect~\citep{Purcell1946}. 
For a cavity with quality factor $Q$ and mode volume $V$, the Purcell effect is characterized by $F_p=\frac{3}{4\pi^2}(\frac{\lambda}{n})^3\frac{Q}{V}$ for a dipole emitter in resonance with the cavity,  placed at the maximum of the electric field, and with proper aligned polarization. $\lambda$ is the wavelength of fundamental mode resonance and $n$ is the material's index of refraction. With the emitter and cavity in resonance, this shortens the radiative lifetime.
While various optical cavities can be used~\citep{Senellart2017}, a particularly useful cavity is the pillar microcavity~\citep{Pelton2002} because the single-photon emission is in a well defined Gaussian mode. 
Since weak cavity coupling reduces the radiative lifetime, decoherence contributions to the emission linewidth are reduced, leading to bandwidths that can approach the spontaneous-emission lifetime-limit, and near unity photon indistinguishability~\citep{Iles2017,Senellart2017}.

Because of the single mode nature of the micropillar cavities, the resonant excitation pump and resonance fluorescence signal are in the same spatial mode. Separating the pump laser from the signal is often achieved through pump-probe cross-polarization, leading to a signal reduction that is at best 50 \%~\citep{Englund2010,Somaschi2016,Unsleber2015,Ding2016,Muller2007}. This reduction in efficiency eliminates quantum applications that require high efficiency (as opposed to brightness)~\citep{Knill2001}. 

It has previously been shown that orthogonal pumping of QDs embedded into planar cavities suppresses scattered laser light~\citep{Muller2007, Muller2008, Jayakumar2013, Thomay2017, Gazzano2018}. Nevertheless, this was limited to planar structures with moderate~\citep{Huber2013} or no Purcell enhancement of the emitter lifetime. Micro-pillar cavities, on the other hand, have much better Purcell enhancement~\citep{Gerard1998, Solomon2001} due to their high Q with a relatively small mode volume, but resonant excitation is limited to cross polarized excitation~\citep{Somaschi2016,Unsleber2015,Ding2016}. Current approaches for orthogonal pumping of micro-pillar cavities are free-space and require cross-polarization, and cannot couple to multiple cavities~\citep{Ates2009}. 

Alternative approaches to in-plane excitation have recently be demonstrated by the Lu group that also remove the 50~\% photon loss associated with cross-polarization. They include using the polarization splitting induced by elliptical cavities\citep{gayral1998}, providing 24~\% efficiency when accounting for the detector efficiency~\citep{lu2019a}, and an alternative approach using a coherent two-color pump source~\citep{lu2019b}.

In this paper, we demonstrate a ridge waveguide-coupled optical cavity architecture where the resonant laser pump and the collected resonance fluorescence are spatially orthogonal. This combines orthogonal waveguide pumping with micro-pillar cavities, allowing for the filter-free off-chip coupling of single photons without the 50\% penalty in source brightness and efficiency present in most current device designs. 
The device design combines the advantages of waveguides~\citep{Monniello2014, Stepanov2015, Javadi2015} with the advantages of cavity QED~\citep{Senellart2017,Gazzano2016}.
 The waveguide enables us to excite several micro-pillar cavities simultaneously, while it significantly reduces laser scattering. 
 We verify our experimental results through simulation, and discuss the limitations of the current design.
 We show that the presented device structure allows for confined cavity modes with a Purcell factor of about 2.5, in-plane guided waveguide modes for excitation, and suppression of unwanted pump laser scattering leading to a filter-free auto-correlation value of $\g_\mathrm{fit}=0^{+0.043}_{-0}$\, where by filter free we mean no spectral, temporal or polarization filtering.
 


\begin{figure}[ht]%
\begin{centering}
\includegraphics[width=1\columnwidth]{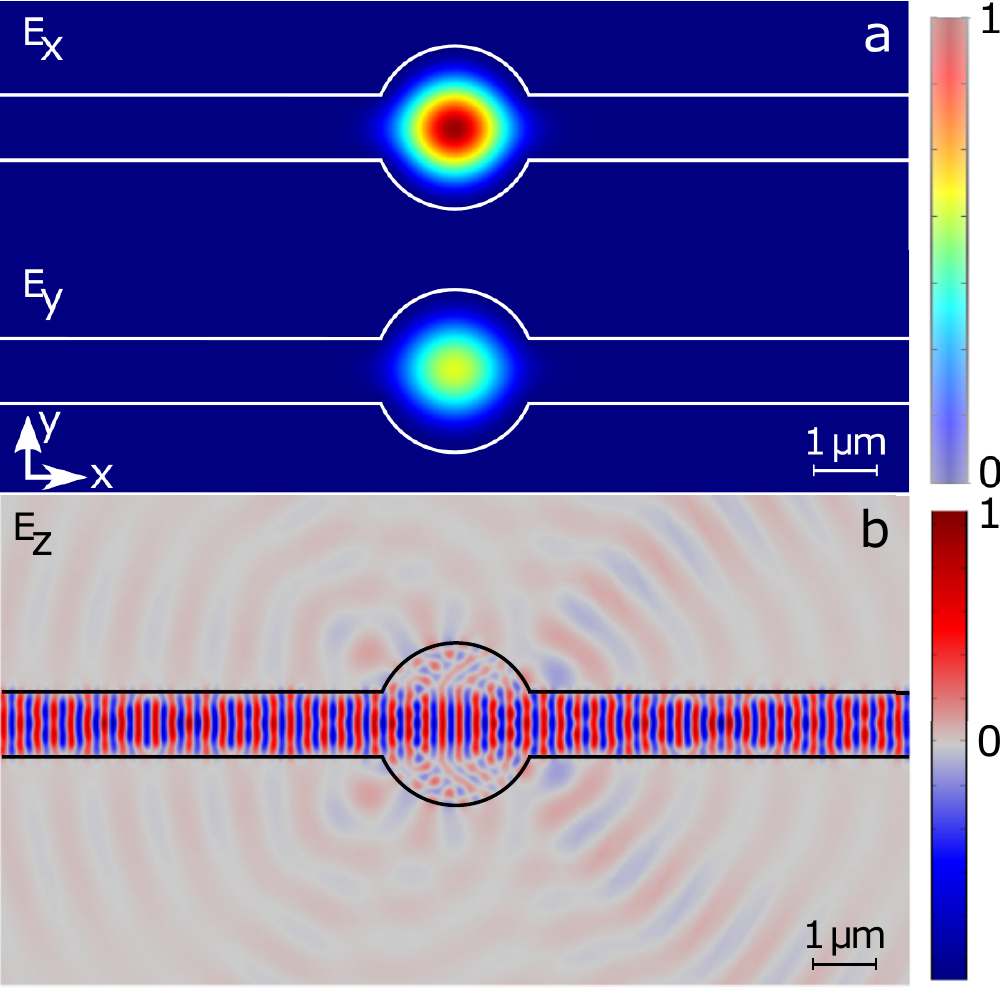}%
\caption{Simulation of the confined modes in the device for a 2.5\un{\mu m} diameter cavity and a 0.95\un{\mu m} waveguide. \textbf{a} intensity ($E_{x(y)}^2$) of the confined modes in the micro-pillar cavity. $E_{x(y)}^2$ is the electric field in the direction along (perpendicular) to the wave guide. These two directions define the two different polarization modes. The polarization modes have a different strength, but have the same spatial extend and their energies overlap within $0.05\un{nm}$. \textbf{b} electric field propagating in the waveguide. No scattering is visible when plotting the intensity, thus we show the electric field distribution for better clarity.} 
\label{fig:comsol}%
\end{centering}
\end{figure}

The device fabrication begins with a distributive-Bragg reflector (DBR) planar microcavity with QDs at the center of a 4-$\lambda$ cavity. (See Supplemental Material and Fig. 2a.)
Our device design minimizes scattering between the waveguide modes, but also maintains confinement in the out-of-plane micro-pillar cavity mode. 
Simulations indicate that the best results are pillar diameters between 2-- and 3--$\mu$m and waveguide widths between 0.55-- and 1.25--$\mu$m, where smaller waveguides increase the cavity confinement and decrease the polarization mode splitting, but increase the scattering at the waveguide-cavity interface. A FDTD simulation of the confined cavity mode and the in-plane waveguide mode can be seen in Fig.~\ref{fig:comsol}.

\begin{figure}[ht]%
\begin{centering}
\includegraphics[]{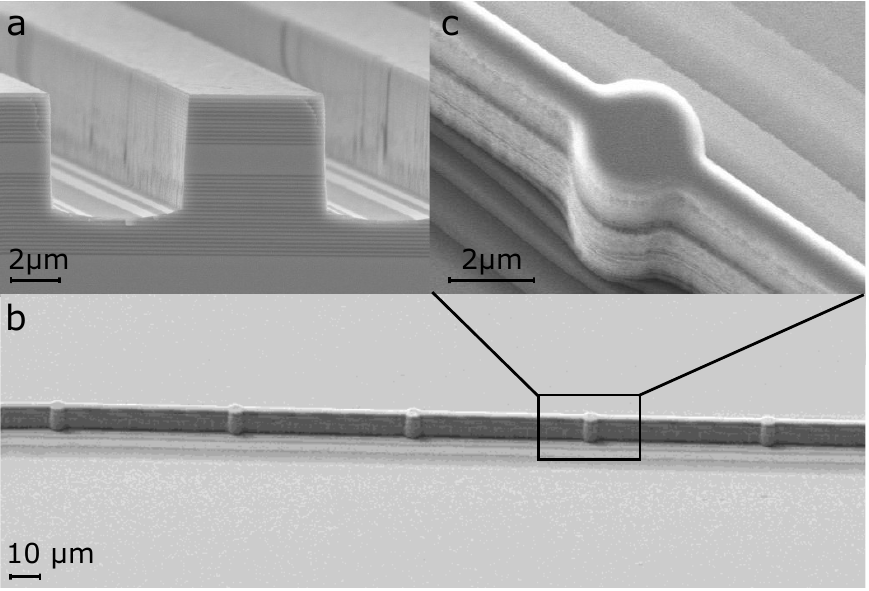}%
\caption{Scanning electrom microscopy (SEM) images of the sample. \textbf{a} Cleaved edge of the sample which is used to couple the laser into the waveguide. The waveguide at the sample edge is 5.5\un{\mu m} wide and adiabatically tapers down to the design width. \textbf{b} The waveguide is connecting micro-pillar cavities, which are used for out-of-plane enhancement of the emission of quantum dots. \textbf{c} zoom on a single micro-pillar cavity. The shown cavity is 2.8\un{\mu m} in diameter and the waveguide is 1.25\un{\mu m} wide.}%
\label{fig:sem}%
\end{centering}
\end{figure}

To suppress residual scattering we planarize the sample with a polymer and cover it with gold, opening circular apertures over the micropillars, allowing outcoupling of the QD emission~\citep{Hopfmann2016} (see Supplemental Material).
The device before planarizing and gold coating is shown in Fig.~\ref{fig:sem}, where Fig.~\ref{fig:sem}~(a) shows the cleaved edge of the device, which is used for coupling of a free space beam. The width of the waveguide then adiabatically tapers down to its design width. The current chip design combines 8 different waveguide widths and pillar-diameters. Fig.~\ref{fig:sem}~(b) shows the waveguide connecting 5 micro-pillar cavities. However, each waveguide connects 25 micro-pillar cavities of the same size along one waveguide, but differ in size for different waveguides. The cavity diameters increase from 2.1\un{\mu m} to 2.8\un{\mu m} and the waveguide width changes from 0.55\un{\mu m} to 1.25\un{\mu m}, both in 0.1\un{\mu m} steps.  Fig.~\ref{fig:sem}~c shows a single micro-pillar cavity.  

\begin{figure}[ht]%
\begin{centering}\emph{}
\includegraphics[width=1\columnwidth]{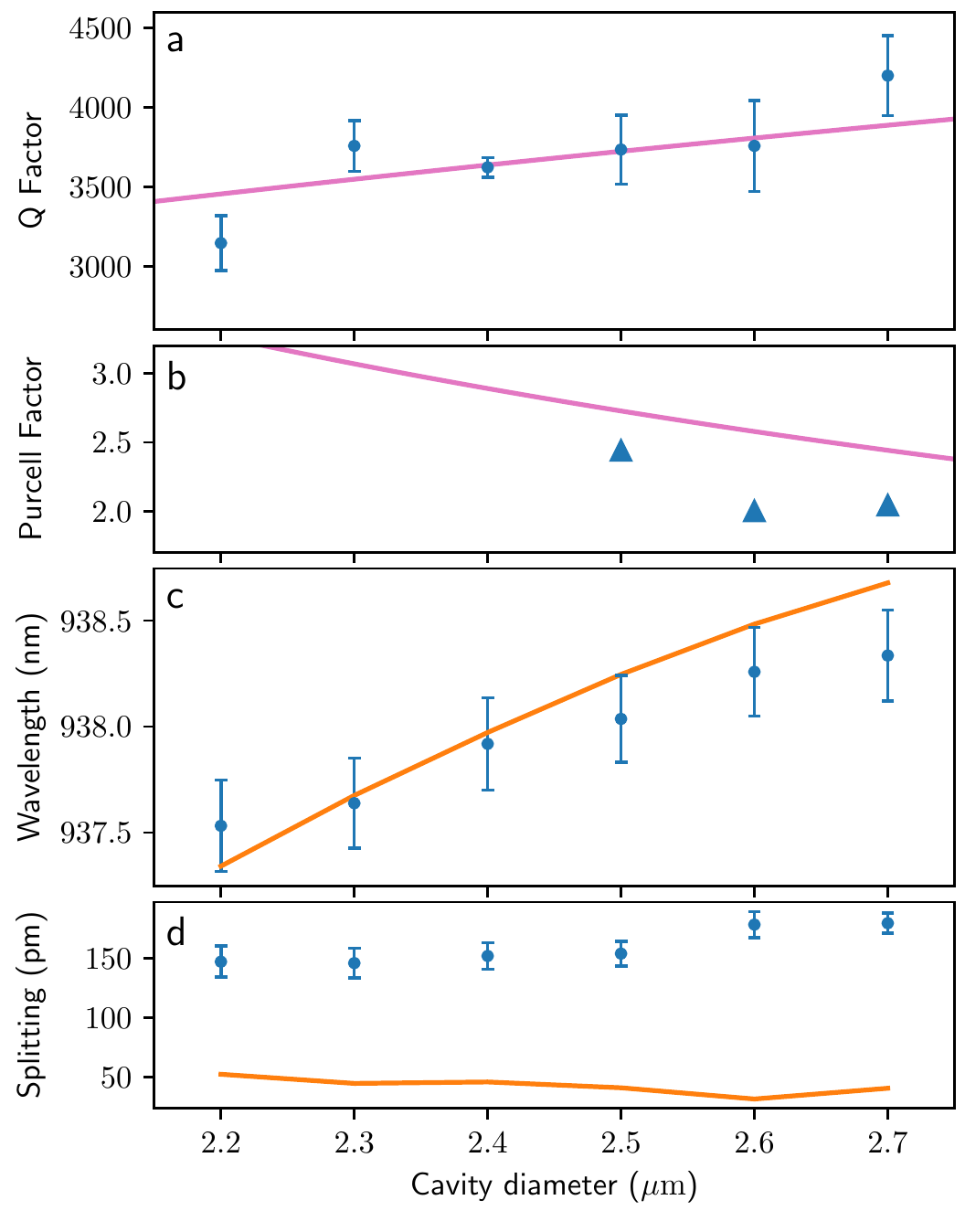}
\caption{ (a) Cavity quality ($Q$) factors and Purcell factors measured before planarization of the sample. Blue dots: mean of measured $Q$ factors, blue triangles: best single measured values for a given cavity diameter. Purple lines: fit of $Q$ factors with sidewall-scattering as a free parameter, and the expected Purcell factor from this fit and a calculated mode volume.  (c) Variation of normal cavity mode wavelength and (d) the normal mode cavity splitting with cavity diameter, where blue dots are again mean values. The orange lines are numerical simulations. Error bars represent one-standard deviation.}%
\label{fig:q}%
\end{centering}
\end{figure}

The $Qs$ of the cavities were measured in photoluminescence, using the QDs as gain medium. The cavity $Qs$ are low enough as to not be significantly affected by any QD absorption. Here, the QDs were pumped above-band with a cw Ti:sapphire laser at 780\un{nm} using a high excitation power density of $P_{\mathrm{pump}}\approx3\times10^3\un{W cm^{-2}}$. The mean of the measured $Q$ factors is plotted  in Fig.~\ref{fig:q}(a). The large error bar comes from the distribution of measured 
$Q$ factors. We assume that this is due to the moderate QD density, where the QD spontanteous emission does not uniformly fill the cavity mode.
To fit the size dependence of the $Q$ factors we used $\frac{1}{Q}=\frac{1}{Q_{planar}}+\frac{1}{Q_{scatt}}$\citep{Rivera1999}, where $Q_{planar}=8350(50)$ is the calculated fundamental mode $Q$ factor of the planar microcavity prior to etching of the micropillars, and $\frac{1}{Q_{scatt}}=\frac{\kappa J^2_0(k_tR)}{R}$ is an explicit function for scattering loss of the micro-pillar of radius $R$ with the Bessel function of the first kind $J_0(k_tR)$, where $k_t=n^2k^2-\beta^2$, with the core refractive index  $n$, the mode propagation constant $\beta$, and the sidewall loss parameter $\kappa$. 
The only free parameter for fitting is $\kappa$ and the fit estimates $\kappa=3.8(2)\times 10^{-10}\un{m}$, comparable to results by others~\citep{Reitzenstein2007,Schneider2016}. The expected Purcell factors are in the range of $2-3$ for the measured $Q$ factors with the mode volume from the electric field distribution from FDTD simulations, see Fig.~\ref{fig:q}(b). Since the {\it Q} values are determined from Fig.~\ref{fig:q}(a), the discrepancy between between the data and simulation is likely due to uncertainty in the FDTD simulations of the electric field distribution originating from finite mesh size. The normal-mode cavity wavelength shifts to shorter wavelength with small cavity diameters is shown in Fig.~\ref{fig:q}(c), reflecting the increased electric field confinement with smaller cavity diameters.
The cavity normal-mode splitting before planarization shown in Fig.~\ref{fig:q}(d) is roughly a factor of three larger than the simulations, indicating either uniform process variations because of the consistency of the offset, or is again, related to the FDTD simulations of the electric field distribution.

\begin{figure}[ht]%
\begin{centering}
\includegraphics[]{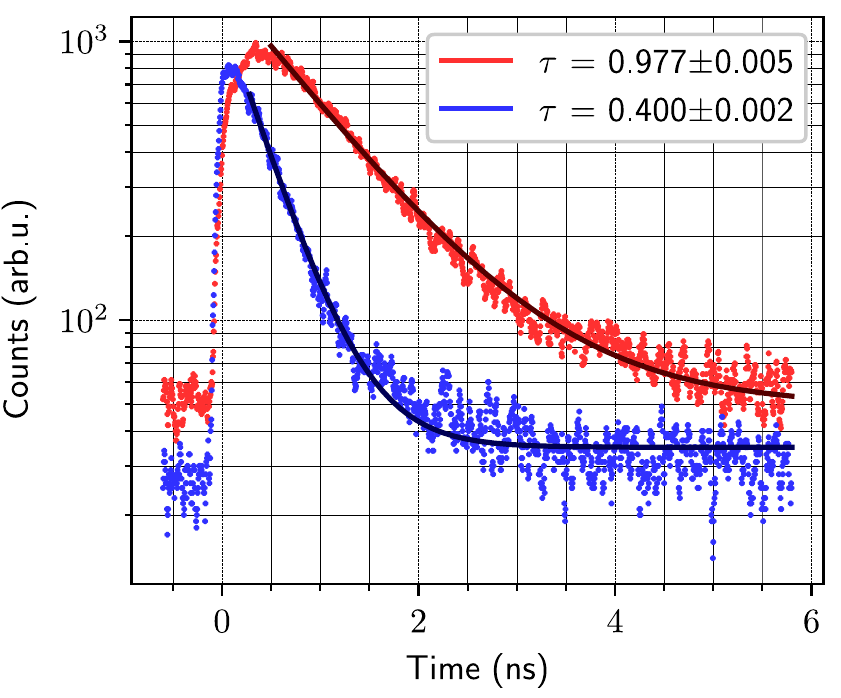}%
\caption{Red: exciton lifetime out of resonance with the cavity mode, blue: exciton lifetime on resonance with the cavity mode. The quantum dot was excited above-band using a 2\un{ps} Ti:sapphire laser at 820\un{nm} with 76\un{MHz} repetition rate. The emission is collected synchronized to the emission laser to extract the lifetime. The excitation power for the red curve was slightly higher than for the blue curve, to measure with comparable count rates. This led to a different rise time of the two curves, probably due to excitation of biexciton-exciton cascades in the off-resonant case. Nevertheless, this is not affecting the measured exciton lifetimes. Uncertainties in the lifetime fit are one standard deviation. }%
\label{fig:lifetimes}%
\end{centering}
\end{figure}

Although the QDs have random emission energy and position, we measure a single-photon lifetime enhancement above 2 for about 10 out of 100 devices at 5 K without tuning. 
An example lifetime measurement is shown in Fig.~\ref{fig:lifetimes}, where we compare the lifetime of an exciton on resonance with an exciton out of resonance to the cavity energy. The Purcell factor is calculated as the ratio of the decay times of an emitter in a cavity and an emitter in bulk Here, approximated by the emitter decay time in the waveguide at the same cavity-resonance wavelength.). Based on the measured lifetimes the Pucrell factor is $F_P=2.44(6)$. 

\begin{figure}[ht]%
\begin{centering}
\includegraphics[]{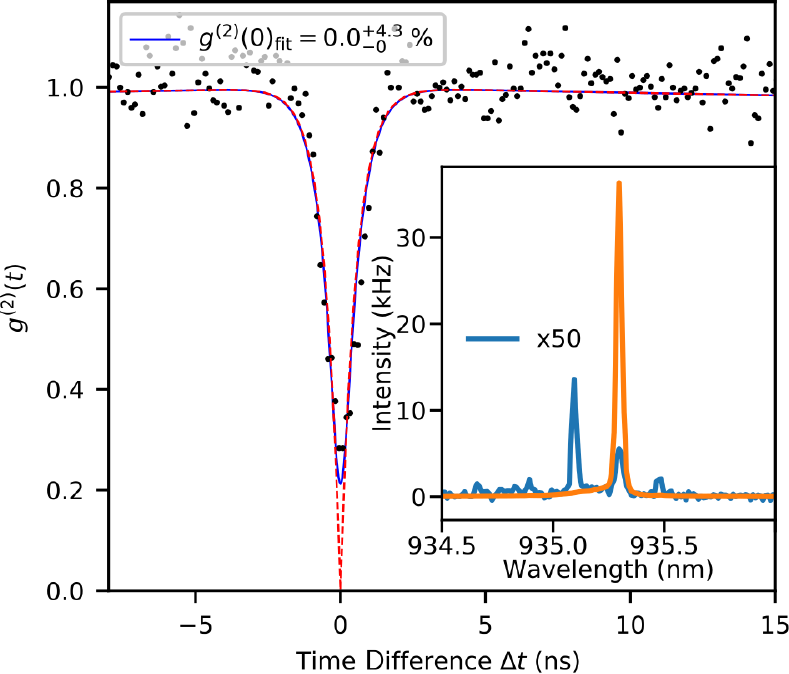}%
\caption{Second-order auto-correlation of photons from a single quantum dot state in the weak excitation resonance fluorescence regime. The fit function is a convolution of the known detector resolution and the expected signal. The blue solid curve is the fit function and the red dashed line is the resulting auto-correlation function for an infinitely fast detector which gives $\g=0.00^{+0.04}_{-0} $ The Rabi frequency is 1 GHz. Uncertainties are one standard deviation.  Inset: Resonance fluorescence when the laser is on resonance (orange) and residual laser scattering (blue) when the laser is detuned by 0.2 nm from the quantum dot resonance with an equivalent Rabi frequency of 6 GHz. The residual scattering signal is displayed a factor of 50 higher than measured, to make the signal visible. }
\label{fig:g2}%
\end{centering}
\end{figure}

To estimate the suppression of the resonant pump laser without filtering, we measure the second-order correlation statistics by exciting a QD state resonantly through the waveguide mode using a tunable cw semiconductor laser. One expects a flat second-order auto-correlation function with a \g ~close to 1 for a Poissonian source, such as an attenuated laser signal, and a dip in the auto-correlation function with a \g=0 for a perfect single photon source.
The measured auto-correlation is shown in Fig.~\ref{fig:g2}.
With no filtering, in resonance fluorescence with a Rabi frequency of $\Omega \approx 1\;GHz$ we find  $\g=0.00^{+0.04}_{-0}$, where the error is calculated from the fit uncertainty. This value of the uncertainty of $\g$ is comparable or better to previously published, where cross-polarization and filters were used~\citep{Englund2010, Unsleber2015,Ding2016,Muller2007,Somaschi2016}. The fit function is a convolution of the known detector response and an exponential function (the single-photon avalanche detectors have a measured detector response of 289(5) ps). To estimate the Rabi frequency, we performed a series of $\g$ measurements and fit the correlation function following Muller \textit{et al.}~\citep{Muller2007}. 

Beyond 1~GHz we cannot characterize the $g^{(2)}(0)$ as we enter the strong light-matter interaction regime. To estimate the laser scattering at high Rabi frequencies, we detune the laser from the QD resonance, see inset in Fig.~\ref{fig:g2}. If we assume this is roughly the resonant value, the estimated laser contribution to the single-photon resonance fluorescence signal from this measurement is $<1~\%$ at a Rabi frequency of 6 GHz. For a Rabi frequency of 6 GHz we measured 4~Mcts/s on the SPAD detectors when the QD is in resonance with the cavity. With a detector quantum efficiency of approximately 0.22 at 930~nm and considering a $10~\%$ counting error due to the detector dead time of 50~ns, the count rate corresponds to approximately 20~Mcts/s on the detector. We note that the large anti-bunching of the device is only present with the metal planarization. Without the metal planarization, the auto-correlation was at best close to 0.5 and in many cases it showed only a very small deviation from 1, as the laser scattering competes with the single-photon resonance fluorescence from the single QD state.


Four parameters are important in the characterization of single photon sources: The source brightness {\it i.e.,} how many useful photons are collected; the source efficiency, {\it i.e.,} the percentage of arbitrary time bins occupied by single photons; suppression of multi-photons, as measured by the second-order correlations ($g^{(2)}(t)$); and the indistinguishability of the quantum light.
In many emerging quantum optics experiments and applications the brightness of the source is critically important to a successful outcome.  For example, boson sampling was simulated using quantum dot single photons \citep{Wang2017}, the source produces 26 million photons per second without normalizing out detector inefficiencies using cross polarization. With our approach this could be boosted by a factor of two, yet in some cases, as in Ref.\citep{Wang2017} where a single photon source is multiplexed, other processes (for instance, Pockell cells) are the limiting factor to useful brightness. For some quantum communications protocols such as BB84, single-photon brightness may provide an appealing advantage to attenuated lasers. For higher order photon correlations this reduces the measurement time by correlation-order squared (e.g. a factor of 4 for $g^{(2)}(t)$),  which  allows the expansion of the number of interacting nodes and photons.
In other applications, such as linear optical quantum computing~\citep{Knill2001} and quantum metrology~\citep{Giovannetti2004,Banaszek2009,Matthews2016,muller2017}, the source efficiency above certain thresholds is critically important while a source brightness is an added benefit.  Single-photon sources require various degrees of multi-photon suppression, but whereas some applications require extremely high indisguishability, others require none.


Our device design has a variety of flexible attributes. The device has partially overlapping cavity modes of orthogonal polarization; thus, the emission can be unpolarized for certain applications such as BB84, or polarized for other applications such as boson sampling. Furthermore, the cavity-mode splitting can be adjusted through processing. However, the device design is not without issues. These include the alignment of the in-plane QD dipole with the waveguide mode for optimum pump light efficiency; and the alignment of the QD with the pillar cavity, which here is not optimized. Both of this issues relate to the classical efficiency of the device; for instance the number of working devices and the pump efficiency, and can be overcome with further engineering.

While the waveguide coupling to the cavity provides efficient in-plane QD resonant excitation, a small component of the QD resonance fluorescence couples back into the waveguide and not into the cavity mode. From simulations, we estimate this to be about 10-15 \% of the total QD emission. Finally, while a count rate of 20 Mcts/s constitutes a bright source, the system is pumped cw and the radiative decay rate is 2.5 GHz. The large difference is due to spectral diffusion induced blinking of the emission. Adding a small amount of nonresonant light can markedly reduce this effect~\citep{metcalfe2010,chen2016}; however, this was not implemented here, to avoid the need for spectral filters.

The presented device  is a first step towards an all integrated single photon source. A future device could divert a small fraction of the light on chip for real-time metrology analysis (the 10-15 \% discussed above), while sending most of the light off chip to be used in an application. 
Such an approach would require low-loss waveguides~\citep{Davanco2017}, on-chip detectors~\citep{faraz2015}, and schemes to measure on-chip indistinguishability and multi-photon suppression~\citep{Thomay2017}. While each presents its own challenges, they are individually useful in various emerging quantum photonics applications.

\bibliographystyle{apsrev4-1}
\bibliography{waveguide_pillars}

\section*{Acknowledgments}
This work is partially funded by the NSF Physics Frontier Center at the Joint Quantum Institute (PFC@JQI).

\section*{Supplemental Material}

A distributed Bragg reflector (DBR) cavity was was grown with 
molecular-beam epitaxy. The cavity consists of 12.5 upper and 20.5 lower DBR pairs of alternating AlAs and GaAs. The $4 \lambda$ cavity/waveguide region contains an active layer of InAs QDs as single photon emitters with a density of about 20\un{\mu m^{-2}}.
The micropillar cavities and ridge waveguides were defined with e-beam lithography and dry etched using an inductively coupled plasma with a $Cl_2-Ar$ gas mixture. 
After etching the DBR planar structures into ridge waveguides, the device was spin coated with a planarization layer of Benzocyclobutene (BCB) and cured at 300~$^\circ$C for one hour in a nitrogen environment.
The sample was then covered with negative photoresist and a laser direct write lithography system was used to produce undercut, circular lift-off resist patterns over the fabricated micropillars exclusively, leaving other areas of the surface exposed.  
Next, 10~nm of Ti and 50~nm of Au were deposited over the sample with an electron-beam evaporator. The metal over the micropillars was lifted-off in a 1-methyl-2-pyrrolidinone (NMP) resist stripper at room temperature for a few minutes. This left the entire BCB surface covered with a metal layer, with circular, micron-scale apertures over the micropillars.

 To excite the QDs, we used a single-mode ring-cavity Ti:sapphire laser and the emission was collected with a high NA objective and was fiber coupled. The light was  then detected with SPAD detectors without passing through a polarizer, spectrometer, Fabry-P\'erot or any other kind of filter.




\end{document}